\documentclass[12pt,a4paper]{article}

\usepackage{amsmath}
\usepackage{amsfonts}
\usepackage{amssymb}
\usepackage{cite}

\begin{document}

\title{{\textsc{From black holes to emergent gravity}}\\}

\author{
\small{\bf Rabin Banerjee}\\
\small{\it S.N.~Bose National Centre for Basic Sciences,}\\
\small{\it Block JD, Sector III, Salt Lake, Kolkata -- 700098, India.}\\[2ex]
\small\texttt{rabin@bose.res.in}\\[5ex]
\rule{0.85\textwidth}{0.6 pt}\\[1ex]
{\it \small{This essay received an Honourable Mention in the Gravity Research Foundation}}\\
{\it \small{2010 Awards for Essays on Gravitation.}}\\[1ex]
\rule{0.85\textwidth}{0.6 pt}\\[3ex]
}

\date{\small\today}

\maketitle

\vspace{0.8 truecm}

\begin{abstract}

\vspace{0.6 truecm}

Many inequivalent approaches to study black holes yield identical results. Any meaningful theory of gravity should explain the origin of this property. Here we show that the basic holomorphic modes characterising the underlying two dimensional conformal symmetry near the horizon bring about this universality. Moreover these modes lead to a law of equipartition of energy for black holes which suggests a statistical origin of gravity. This emergent nature of gravity is further bolstered by showing the equivalence of entropy with the action $\left(S=-\frac{i}{\hbar}\,I\right)$ and expressing the generalised Smarr formula for mass as a thermodynamic relation, $S=\frac{E}{2T}$ where $S$, $E$ and $T$ are the entropy, energy and temperature, respectively, of a black hole.

\end{abstract}

\pagebreak

\section{Introduction}
\label{Intro}

In the absence of a quantum theory of gravity, semiclassical approaches have become increasingly popular and are widely used. Of particular interest is the application of these ideas to black hole thermodynamics. Black holes are to gravity what atoms were to atomic physics so that their study is expected to provide crucial information regarding gravity just as the study of atoms was the pathway to atomic physics. An intriguing aspect is that identical results for black hole thermodynamics are obtained by quite contrasting semiclassical methods. It is both remarkable as well as ill understood.

Now another consequence of the repeated failure to quantise gravity has led to a parallel development where the interpretation of gravity as a fundamental force is abandoned and it is regarded as an emergent phenomenon like thermodynamics or hydrodynamics \cite{Jacobson:1995ab, Padmanabhan:2009vy, Verlinde:2010hp, Banerjee:2010yd}. Hence the issue of quantisation of gravity becomes inconsequential. A natural question that arises in this context is the role of semiclassical methods in interpreting gravity as an emergent phenomenon. Unless this is clarified, a self-consistent  holistic view of either black holes or gravity is not possible.

In this essay we address these issues. The results to be used were recently obtained by us \cite{Banerjee:2010yd, Banerjee:2008sn, Banerjee:2009wb, Banerjee:2009pf, Banerjee:2010ma, Banerjee:2007qs, Banerjee:2008wq}. These findings are connected to present a unified picture.

The origin of the universality of results lies in the fact that, to an observer outside a black hole, the near horizon region is effectively described by a free two dimensional conformal theory \cite{Carlip:1998wz}. Transverse modes, interaction and mass terms are redshifted away relative to degrees of freedom in the ($r$-$t$) plane. As we shall show this two dimensional effective theory is able to simultaneously accommodate some of the most widely used semiclassical techniques -- anomaly based \cite{Christensen:1977jc, Robinson:2005pd, Iso:2006wa, Banerjee:2007uc, Banerjee:2008az, Banerjee:2007qs}, tunneling based \cite{Srinivasan:1998ty, Parikh:1999mf, Banerjee:2008cf} and global embedding \cite{Deser:1998xb, Banerjee:2010ma} -- to analyse black hole thermodynamics. The essential thread running through these seemingly disparate approaches is the structure of the holomorphic modes characterising the left moving and right moving solutions of the field equation.

The same modes are used to concretely develop a statistical formulation highlighting gravity as an emergent phenomenon. First, the law of equipartition of energy is derived for black holes. Using this, the standard statistical definition of entropy is identified with the action. Consequently, extremisation of the action that yields Einstein's equations becomes equivalent with extremisation of entropy. Our analysis naturally leads to a thermodynamical definition of energy ($E=2\,TS$). This is useful since energy is one of the most conceptually challenging and diversely defined variables in general relativity. This relation is also shown to reproduce the generalised Smarr formula for mass \cite{Smarr:1972kt}.

\section{Mode functions and black hole thermodynamics}
\label{ModeThermodynamics}

Let us now concentrate on the effective two dimensional theory near the event horizon. The left (L) and right (R) moving (holomorphic) modes are obtained by solving the appropriate field equation using the geometrical (WKB) ansatz. The modes defined inside and outside the horizon are related by the transformations \cite{Banerjee:2008sn, Banerjee:2009wb, Banerjee:2009pf},
\begin{eqnarray}
\label{modes}
\phi_{\text{in}}^{\text{(R)}}&=& e^{-\frac{i}{\hbar}\omega \,u_{\text{in}}}=e^{-\frac{\pi \omega}{\hbar\kappa}}~e^{-\frac{i}{\hbar}\omega \,u_{\text{out}}}=e^{-\frac{\pi \omega}{\hbar\kappa}}~\phi_{\text{out}}^{\text{(R)}}\nonumber\\
\phi_{\text{in}}^{\text{(L)}}&=& e^{-\frac{i}{\hbar}\omega \,v_{\text{in}}}=e^{-\frac{i}{\hbar}\omega \,v_{\text{out}}}=\phi_{\text{out}}^{\text{(L)}}
\end{eqnarray}
where $\omega$ is the energy of the particle as measured by an asymptotic observer, $\kappa$ is the surface gravity and $u=t-r_\star$, $v=t+r_\star$ are the null tortoise coordinates $\left(r_\star =\int\frac{dr}{F(r)}\right)$ corresponding to the $2-d$ metric $ds^2=-F(r)dt^2+\frac{dr^2}{F(r)}$. In this convention the L (R) moving modes are ingoing (outgoing). The results obtained in the tunneling formalism are now easily reproduced. Considering the modes inside the horizon, the left moving modes travel towards the centre of the black hole and their probability to go inside, as measured by an external observer, is expected to be unity. This is easily verified by using \eqref{modes}, $P^{\text{(L)}}=\left\vert \phi_{\text{in}}^{\text{(L)}} \right\vert^2=\left\vert \phi_{\text{out}}^{\text{(L)}} \right\vert^2=1$. Note that $ \phi_{\text{in}}^{\text{(L)}}$ is recast in terms of $ \phi_{\text{out}}^{\text{(L)}}$ since measurements are done by an external observer. This shows that the left moving (ingoing) mode is trapped inside the black hole, as expected. The right moving mode, on the other hand, tunnels through the horizon with a finite probability
\begin{eqnarray}
\label{outProbab}
P^{\text{(R)}}=\left\vert \phi_{\text{in}}^{\text{(R)}} \right\vert^2=\left\vert e^{-\frac{\pi \omega}{\hbar\kappa}}\,\phi^{\text{(R)}}_{\text{out}}\right\vert^2 =e^{-\frac{2\pi \omega}{\hbar\kappa}},
\end{eqnarray}
as measured by an external observer. Exploiting the principle of detailed balance $P^{\text{(R)}}=e^{-\frac{\omega}{T}}~P^{\text{(L)}}$ immediately leads to the Hawking temperature $T=\frac{\hbar\kappa}{2\pi}$. Usual tunneling based methods \cite{Srinivasan:1998ty, Parikh:1999mf} also yield this result but fail to obtain the black body spectrum. This glaring omission is filled in the present approach by computing the density matrix using \eqref{modes} and then tracing out the irrelevant ingoing (L) modes \cite{Banerjee:2009wb}.

Let us next consider anomaly based methods. The stress tensor near an evaporating black hole is calculated \cite{Davies:1977} by using precisely the mode functions $\phi \sim e^{-i\,\omega\,u}, e^{-i\,\omega\,v}$ given in \eqref{modes}. Its explicit form is split into a traceful and traceless part $\left(\Theta^\mu_{~\mu}=0\right)$,
\begin{eqnarray}
\label{EMtensor}
T_{\mu\nu}=\frac{R}{48\pi}\,g_{\mu\nu}+\Theta_{\mu\nu}~.
\end{eqnarray}
The stress tensor is conserved $\nabla^\mu T_{\mu\nu}=0$ but has a non-vanishing trace $T^\mu_{~\mu}=\frac{R}{24 \pi}$ since, simultaneously, diffeomorphism and conformal invariances cannot be preserved at the quantum level. This trace anomaly incidentally leads to the Hawking flux as demonstrated long back in \cite{Christensen:1977jc}.

A more recent approach, based on the diffeomorphism anomaly \cite{Robinson:2005pd, Iso:2006wa, Banerjee:2007qs, Banerjee:2007uc}, is easily connected to this discussion. Since the ingoing (L) modes are trapped within the black hole, the theory near the horizon is interpreted as chiral, having only one sector (R) of the modes. Using a simple algebra based on the left-right symmetry the diffeomorphism anomaly $\nabla^\mu T_{\mu\nu}^{\text{(R)}} = \frac{1}{96\pi} \epsilon_{\nu\mu}\nabla^\mu R$ follows immediately from \eqref{EMtensor} \cite{Banerjee:2008sn}. As shown in \cite{Banerjee:2008wq} this equation is solved subject to a boundary condition that corresponds to implementing the Unruh vacuum. The resulting flux agrees with the expression found by integrating the Hawking black body spectrum.

Yet another way of discussing black hole thermodynamics is to use the correspondance between Unruh and Hawking effects. In this case the $n$-dimensional curved space is embedded in $m \,(>n)$ dimensional flat space \cite{Deser:1998xb}. This global embedding is quite complicated for an arbitrary black hole metric. As shown recently \cite{Banerjee:2010ma}, restricting to the effective two dimensional near horizon theory simplifies the calculations non-trivially. The crucial point is that, after embedding in flat space, the relations \eqref{modes} between the `in' and `out' modes is still preserved, although the individual form of the modes might differ from the curved space result. The only difference is that the asymptotic energy `$\omega$' in \eqref{modes} is replaced by the local energy given by the Tolman relation $E(r)=\omega\, F(r)$. Following similar steps beginning from \eqref{outProbab} it is possible to derive the black body spectrum with the Unruh temperature $T_U=\frac{\hbar \kappa}{2 \pi \sqrt{F(r)}}$ \cite{Roy:2009vy}. The Hawking temperature is easily deduced from this expression by using $T=\sqrt{F(r)}T_U=\dfrac{\hbar \kappa}{2 \pi}$.

\section{Mode functions and statistical origin of gravity}
\label{ModeEmGrav}

Recently there has been a growing consensus that gravity is not a fundamental force, rather it is an emergent phenomenon \cite{Jacobson:1995ab, Padmanabhan:2009vy, Verlinde:2010hp, Banerjee:2010yd}. The fundamental role of gravity is replaced by thermodynamical interpretations. However, understanding the entropic or thermodynamic origin of gravity is far from complete since the arguments are more heuristic than concrete. We now show that the relations \eqref{modes} play a crucial role in revealing the statistical origin of gravity.

First, we give a simple derivation of the law of equipartition of energy of black holes. The average value of the energy of a particle, measured from outside, is given by \cite{Banerjee:2009pf},
\begin{eqnarray}
\label{avEnergy}
\langle \omega \rangle = \dfrac{\displaystyle\int_0^\infty d\omega \, \omega \, e^{-\frac{2 \pi \omega}{\hbar\kappa}}}{\displaystyle\int_0^\infty d\omega \, e^{-\frac{2 \pi \omega}{\hbar\kappa}}}=\dfrac{\hbar\kappa}{2\pi}=T
\end{eqnarray}
where we have used \eqref{outProbab} and $T$ is the black hole temperature. Now the effective two-dimensional curved metric can always be embedded in a flat space having exactly two space-like coordinates \cite{Banerjee:2010ma}. Recall that this type of embedding was discussed at the end of Section \ref{ModeThermodynamics} in the treatment of Unruh radiation vis-a-vis Hawking radiation. Hence, associating two degrees of freedom to a particle, we find that each degree must have an energy equal to $\frac{T}{2}$. This is just the equipartition law of energy.

Imagine now that the system is composed of $N$ number of bits corresponding to $N$ degrees of freedom, in which all information is stored, so that $N$ is proportional to the entropy, $N=N_0\,S$. From the equipartition law the total energy of the system is $E=\frac{1}{2}\,NT$. Eliminating $N$, we obtain $S=\frac{2E}{N_0T}$. To fix $N_0$ let us consider the simplest example, the Schwarzschild black hole with $S=\frac{4\pi M^2}{\hbar}$, $E=M$, $T=\frac{\hbar}{8\pi M}$. This yields $N_0=4$. With this normalisation we obtain,
\begin{eqnarray}
\label{TheReln}
S=\frac{E}{2T}.
\end{eqnarray}
Now in statistical physics, entropy is defined as $S=\ln Z + \frac{E}{T}$. Expressing the partition function as $Z=e^{\frac{i}{\hbar}\,I}$ and exploiting \eqref{TheReln} immediately yields,
\begin{eqnarray}
\label{ActionEntropy}
S=-\frac{i}{\hbar}\,I.
\end{eqnarray}
Consequently the gravity action gets identified with the entropy. Furthermore, using \eqref{TheReln}, \eqref{ActionEntropy} and the explicit form of the action, it is possible to prove \cite{Banerjee:2010yd} that the energy $E$ corresponds to Komar's definition \cite{Komar:1958wp}.

The relation \eqref{TheReln}, which was also considered in \cite{Banerjee:2010yd, Padmanabhan:2003pk}, has a general validity as may be explicitly verified by taking various black hole solutions \cite{Banerjee:2010yd}. Incidentally it is also possible to interpret \eqref{TheReln} as defining the energy. For instance, in the Kerr-Newman example, the Komar energy is not known in a closed form. Here if we use the known expressions for $S$ and $T$, the Komar energy comes out as,
\begin{eqnarray}
\label{KNKomar}
E=M-QV-2\,J\Omega
\end{eqnarray}
where the symbols have their usual meaning. If $E=2\,TS$ is re-introduced in \eqref{KNKomar}, we exactly reproduce the generalised Smarr formula for mass \cite{Banerjee:2010yd, Smarr:1972kt}. The thermodynamic origin of this formula is thereby illuminated.

\section{Concluding remarks}
\label{conclu}

Semiclassical methods, along with other approaches like loop quantum gravity and string theory, form the basis for studying black hole thermodynamics. It is reassuring to note that, in the supposed absence of a quantum theory of gravity, the same semiclassical methods are useful to study gravity as an emergent phenomenon. The unification of the semiclassical methods, eventually leading to a concrete statistical origin of gravity shown here, could further illuminate this phenomenon.


\begin{thebibliography}{99}

\bibitem{Jacobson:1995ab}
  T.~Jacobson,
  Phys.\ Rev.\ Lett.\  {\bf 75}, 1260 (1995)
  [arXiv:gr-qc/9504004].
\bibitem{Padmanabhan:2009vy}
  T.~Padmanabhan,
  arXiv:0911.5004 [gr-qc] and references therein.
\bibitem{Verlinde:2010hp}
  E.~P.~Verlinde,
  arXiv:1001.0785 [hep-th].
\bibitem{Banerjee:2010yd}
  R.~Banerjee and B.~R.~Majhi,
  arXiv:1003.2312 [gr-qc].
\bibitem{Banerjee:2008sn}
  R.~Banerjee and B.~R.~Majhi,
  Phys.\ Rev.\  D {\bf 79}, 064024 (2009)
  [arXiv:0812.0497 [hep-th]].
\bibitem{Banerjee:2009wb}
  R.~Banerjee and B.~R.~Majhi,
  Phys.\ Lett.\  B {\bf 675}, 243 (2009)
  [arXiv:0903.0250 [hep-th]].
\bibitem{Banerjee:2009pf}
  R.~Banerjee, B.~R.~Majhi and E.~C.~Vagenas,
  Phys.\ Lett.\  B {\bf 686}, 279 (2010)
  [arXiv:0907.4271 [hep-th]].
\bibitem{Banerjee:2010ma}
  R.~Banerjee and B.~R.~Majhi,
  arXiv:1002.0985 [gr-qc].
\bibitem{Banerjee:2007qs}
  R.~Banerjee and S.~Kulkarni,
  Phys.\ Rev.\  D {\bf 77}, 024018 (2008)
  [arXiv:0707.2449 [hep-th]].
\bibitem{Banerjee:2008wq}
  R.~Banerjee and S.~Kulkarni,
  Phys.\ Rev.\  D {\bf 79}, 084035 (2009)
  [arXiv:0810.5683 [hep-th]].
\bibitem{Carlip:1998wz}
  S.~Carlip,
  Phys.\ Rev.\ Lett.\  {\bf 82}, 2828 (1999)
  [arXiv:hep-th/9812013].
\bibitem{Christensen:1977jc}
  S.~M.~Christensen and S.~A.~Fulling,
  Phys.\ Rev.\  D {\bf 15}, 2088 (1977).
\bibitem{Robinson:2005pd}
  S.~P.~Robinson and F.~Wilczek,
  Phys.\ Rev.\ Lett.\  {\bf 95}, 011303 (2005)
  [arXiv:gr-qc/0502074].
\bibitem{Iso:2006wa}
  S.~Iso, H.~Umetsu and F.~Wilczek,
  Phys.\ Rev.\ Lett.\  {\bf 96}, 151302 (2006)
  [arXiv:hep-th/0602146].
\bibitem{Banerjee:2007uc}
  R.~Banerjee and S.~Kulkarni,
  Phys.\ Lett.\  B {\bf 659}, 827 (2008)
  [arXiv:0709.3916 [hep-th]].
\bibitem{Banerjee:2008az}
  R.~Banerjee,
  Int.\ J.\ Mod.\ Phys.\  D {\bf 17}, 2539 (2009)
  [arXiv:0807.4637 [hep-th]].
\bibitem{Srinivasan:1998ty}
  K.~Srinivasan and T.~Padmanabhan,
  Phys.\ Rev.\  D {\bf 60}, 024007 (1999)
  [arXiv:gr-qc/9812028].
\bibitem{Parikh:1999mf}
  M.~K.~Parikh and F.~Wilczek,
  Phys.\ Rev.\ Lett.\  {\bf 85}, 5042 (2000)
  [arXiv:hep-th/9907001].
\bibitem{Banerjee:2008cf}
  R.~Banerjee and B.~R.~Majhi,
  JHEP {\bf 0806}, 095 (2008)
  [arXiv:0805.2220 [hep-th]].
\bibitem{Deser:1998xb}
  S.~Deser and O.~Levin,
  Phys.\ Rev.\  D {\bf 59}, 064004 (1999)
  [arXiv:hep-th/9809159].
\bibitem{Smarr:1972kt}
  L.~Smarr,
  Phys.\ Rev.\ Lett.\  {\bf 30}, 71 (1973)
  [Erratum-ibid.\  {\bf 30}, 521 (1973)].
\bibitem{Davies:1977}
  P.~C.~W.~Davies and S.~A.~Fulling,
  Proc.\ Roy.\ Soc.\ Lond.\  A {\bf 354} (1977) 59.\\
  S.~A.~Fulling,
  Gen.\ Rel.\ Grav.\  {\bf 18} (1986) 609.
\bibitem{Roy:2009vy}
  D.~Roy,
  Phys.\ Lett.\  B {\bf 681}, 185 (2009)
  [arXiv:0908.3149 [hep-th]].
\bibitem{Komar:1958wp}
  A.~Komar,
  Phys.\ Rev.\  {\bf 113}, 934 (1959).
\bibitem{Padmanabhan:2003pk}
  T.~Padmanabhan,
  Class.\ Quant.\ Grav.\  {\bf 21}, 4485 (2004)
  [arXiv:gr-qc/0308070].

\end{thebibliography}
\end{document}